# Parallel Paradigms in Modern HPC: A Comparative Analysis of MPI, OpenMP, and CUDA

Nizar ALHafez*, nizar.alhafez@hiast.edu.sy
Ahmad Kurdi*, ahmad.kurdi@hiast.edu.sy
*Department of Informatics Engineering, Higher Institute for Applied Sciences and Technology (HIAST),
Damascus, Syria

*Abstract*—This paper presents a comprehensive comparison of three dominant parallel programming models in High Performance Computing (HPC): Message Passing Interface (MPI), Open Multi-Processing (OpenMP), and Compute Unified Device Architecture (CUDA). As computational demands grow exponentially across scientific and industrial domains, selecting optimal programming approaches for modern heterogeneous HPC architectures has become increasingly critical. We systematically analyze these models through multiple dimensions: architectural foundations, performance characteristics, domain-specific suitability, programming complexity, and recent advancements. We examine each model's strengths, weaknesses, and optimization techniques. Our investigation demonstrates that MPI excels in distributed memory environments with near-linear scalability for communication-intensive applications but faces communication overhead challenges, while OpenMP provides superior performance and usability in shared-memory systems and loop-centric task through its directive-based approach but is also limited by shared memory contention. CUDA dominates in data-parallel tasks with substantial performance gains for GPU-compatible workloads but is limited to NVIDIA GPUs and requires specialized expertise. Performance evaluations across scientific simulations, machine learning, and data analytics reveal that hybrid approaches—combining two or more models—often yield optimal results in heterogeneous computing environments. The paper further explores implementation challenges, optimization best practices, and emerging trends including performance portability frameworks, task-based programming models, and HPC-Big Data convergence. This research provides a foundation for developers and researchers to make informed decisions when selecting programming models for modern HPC applications, highlighting that the optimal choice depends on application characteristics, target architecture, and development constraints rather than a one-size-fits-all solution.

*Keywords*—High-Performance Computing (HPC), Parallel Programming, Message Passing Interface (MPI), Open Multi-Processing (OpenMP), Compute Unified Device Architecture (CUDA).

## I. Introduction

The exponential growth in data volume and computational complexity has driven the adoption of High-Performance Computing (HPC) across scientific, industrial, and commercial domains. The increasing complexity of scientific problems and the ever-growing volume of data in modern research necessitate computational resources that can deliver unprecedented levels of performance [1]. HPC systems have emerged as indispensable tools to tackle these computationally intensive tasks, enabling advancements across a wide spectrum of disciplines, including climate modeling, molecular dynamics (MD), and artificial intelligence (AI). The ability of HPC to handle intricate simulations and analyze massive datasets has become pivotal in driving scientific discovery and technological innovation. The continuous surge in data generation and the escalating complexity of computational models further amplify the critical need for efficient HPC solutions to process and interpret this information effectively.

At the heart of harnessing the power of HPC lies the effective utilization of parallel programming models. These models provide the frameworks and tools necessary to divide complex computational problems into smaller tasks that can be executed concurrently across multiple processors or computing units. Modern HPC architectures, characterized by multi-core CPUs, GPUs, and distributed clusters, require tailored APIs to optimize performance. Among the most prominent parallel programming models in the HPC landscape are the Message Passing Interface (MPI), Open Multi-Processing (OpenMP), and Compute Unified Device Architecture (CUDA) [1][2][3][4]. These models represent the fundamental approaches to parallelizing computations for HPC applications, each with its own characteristics and suitability for different types of hardware architectures and problem structures.

MPI enables distributed memory parallelism through message passing [4][5], OpenMP simplifies shared-memory multithreading via compiler directives [5], and CUDA unlocks massive parallelism on GPUs [3]. Each of these models targets distinct system paradigms: distributed memory (MPI), shared memory (OpenMP), and GPU acceleration (CUDA). A thorough understanding of their principles, recent advancements, and limitations is essential for researchers aiming to leverage the full potential of HPC resources.

This paper synthesizes findings from recent studies to compare these models in terms of performance, scalability, programming complexity, and application suitability. By analyzing abstraction levels, performance metrics, and programmer experiences, we aim to guide developers and educators in selecting appropriate tools for HPC applications. We will delve into a comparative study of these three primary parallel programming models within the context of HPC, drawing upon recent research to provide a comprehensive overview of their strengths, weaknesses, and evolving roles in addressing the ever-increasing demands of computational science.

The remainder of this paper is organized as follows:

Section 2 provides an in-depth analysis of the three programming models—MPI, OpenMP, and CUDA—highlighting their architectural foundations, core functionalities, recent advancements, and inherent strengths and limitations. Section 3 presents a comparative performance evaluation across diverse HPC application domains, including scalability benchmarks, scientific simulations, and machine learning workloads, to elucidate the suitability of each model under varying computational demands. Section 4 explores hybrid programming paradigms, such as MPI+OpenMP, MPI+CUDA, and unified MPI+OpenMP+CUDA frameworks, demonstrating their synergistic potential in heterogeneous HPC environments. Section 5 discusses critical challenges in scaling these models on modern architectures, alongside best practices for optimization, fault tolerance, and energy efficiency. Section 6 examines emerging trends, including performance portability frameworks, task-based programming, and the convergence of HPC with big data ecosystems, while Section 7 concludes with insights into the future trajectory of parallel programming in HPC.

## II. Individual Deep Dive

### A. Message Passing Interface (MPI)

The Message Passing Interface (MPI) stands as a cornerstone for parallel programming in distributed memory systems [1]. Its core principle revolves around enabling communication and data exchange between independent processes, each residing in its own memory space, often across different nodes in an HPC cluster. MPI is a standardized library for distributed memory systems, enabling communication between processes across cluster nodes via message passing.

MPI facilitates this interaction through a rich set of functions that support both point-to-point communication (e.g., MPI_Send, MPI_Recv) between pairs of processes and collective communication (e.g., MPI_Broadcast, MPI_Reduce) involving groups of processes [6]. This capability makes MPI particularly crucial for applications that require significant inter-node communication, a common scenario in large-scale scientific simulations and data-intensive computations. MPI's strength lies in scalability for large-scale distributed applications, though its low-level nature increases code complexity [3].

Recent advancements in MPI standards and their implementations have been driven by the need to enhance performance and scalability on modern HPC clusters, which are increasingly characterized by high core counts, multi-threading within nodes, and the integration of accelerators like GPUs. The latest iteration of the standard, MPI-4, introduces several key features designed to address these evolving architectural trends [7]. Partitioned communications, for instance, offer increased flexibility and enable the overlapping of communication and computation, a crucial optimization technique for reducing overall execution time. Furthermore, the concept of Sessions, along with the Process Management Interface for Exascale (PMIx), brings a new level of isolation and dynamic resource allocation to MPI environments, allowing for more efficient management of complex HPC workloads.

Optimized MPI libraries, such as the Intel MPI Library, have demonstrated significant performance improvements in various HPC workloads by leveraging advanced network offloads and supporting the latest hardware features [7]. Similarly, MVAPICH2-GDR provides an optimized MPI implementation specifically for clusters equipped with NVIDIA GPUs, which is particularly beneficial for emerging deep learning applications that heavily rely on GPU acceleration. These advancements underscore the continuous effort to adapt and enhance MPI to meet the demands of contemporary HPC systems.

Despite its widespread adoption and ongoing development, MPI applications face several challenges when deployed at large scales. One significant hurdle is the potential for communication overhead to limit scalability, especially in strong scaling scenarios where the problem size remains fixed while the number of processors increases. Managing data locality and minimizing communication latency become paramount for achieving optimal performance in such environments. Additionally, MPI relies on manual checkpointing and restart mechanisms for fault tolerance, which are becoming increasingly impractical for exascale systems [4].

To mitigate these challenges, hybrid programming models that combine MPI with shared memory parallelization approaches like OpenMP or accelerator-based programming models like CUDA are frequently employed. Recent research also highlights best practices for running MPI workloads on cloud platforms, including the use of compute-optimized virtual machines, compact placement policies to reduce inter-node latency, and careful tuning of MPI libraries for specific hardware configurations [7].

Key strengths of MPI include:
1) Scalability across distributed nodes.
2) Flexibility in handling dynamic and irregular data structures.
3) Support for fault tolerance through checkpointing.
4) Portability and popularity, as it is widely implemented on clusters and supercomputers.
5) Hybridity, as it can be combined with multithreading APIs like OpenMP or Pthreads for intra-node parallelism.

Key weaknesses include:

1) High programming complexity due to explicit communication management, handling of data distribution, and synchronization.
2) Performance bottlenecks in shared-memory systems compared to OpenMP.

*B. Open Multi-Processing (OpenMP)*

OpenMP presents a directive-based Application Program Interface (API) that facilitates parallel programming on shared memory architectures, most commonly found in multi-core CPUs [5]. It operates based on a fork-join model, where a master thread spawns a team of threads to execute parallel regions of code, and these threads subsequently join the master thread upon completion. OpenMP's ease of use, through the insertion of compiler directives, has made it a popular choice for parallelizing applications on multi-core processors.

OpenMP provides directive-based parallelism for shared memory systems, simplifying multithreaded programming on multi-core CPUs [3][4][5]. It uses pragmas to manage thread creation, workload distribution, and synchronization. This high-level abstraction minimizes code modifications and has been rated as the easiest API to use by students due to its intuitive syntax. The simplicity of OpenMP allows for rapid parallelization of sequential code.

Recent advancements have significantly expanded the scope of OpenMP, most notably with the introduction of support for offloading computations to accelerators like GPUs [4][7]. Newer versions of the standard, particularly 5.0 and beyond, have introduced features such as loop directives and meta directives, which offer more fine-grained control over parallel execution and data management on heterogeneous systems. These advancements aim to improve both the performance achieved on these systems and the productivity of developers. The release of OpenMP 6.0 further underscores this trend, with enhancements focused on simplifying task programming, providing better support for Fortran array syntax and the latest versions of C and C++, and offering finer management of allocatable variables and memory [7]. This evolution indicates that OpenMP is increasingly becoming a versatile option for programming heterogeneous HPC systems, providing a higher level of abstraction compared to lower-level models like CUDA.

While OpenMP simplifies parallel programming on shared memory systems and is expanding its reach to accelerators, achieving scalability on very large-scale systems, especially those with distributed memory, presents challenges. In such scenarios, OpenMP is often combined with MPI in hybrid programming models to leverage the strengths of both approaches. Effective optimization with OpenMP often involves careful management of data locality to ensure that threads access data efficiently from the cache, as well as strategies to avoid false sharing, a performance-degrading phenomenon that can occur when threads access different data items within the same cache line. Techniques such as loop transformations and the strategic use of OpenMP directives play a crucial role in achieving high performance.

Key strengths of OpenMP include:
1) Low overhead for thread management.
2) Ease of integration with existing code via compiler directives.
3) High-level directives that minimize code modifications.
4) Efficient utilization of multi-core CPUs.
5) Simplicity in parallelizing loops and task-based workflows.

Key weaknesses include:
6) Limited to shared-memory systems without extensions.
7) Challenges in load balancing for irregular workloads.
8) Limited native fault tolerance; failed threads often necessitate full application restart.
9) Bottlenecks in fine-grained synchronization.

*C. Compute Unified Device Architecture (CUDA)*

Compute Unified Device Architecture (CUDA) is a parallel computing platform and programming model developed by NVIDIA, specifically designed for leveraging the massive parallelism offered by their Graphics Processing Units (GPUs) [1][5]. CUDA enables programmers to write kernels, which are functions that execute in parallel across thousands of GPU cores, making it exceptionally well-suited for computationally intensive tasks that exhibit high degrees of data parallelism.

CUDA provides a hierarchical parallelism model with grids of thread blocks executing parallel tasks, leveraging GPU memory hierarchies (global, shared, registers) [3]. It extends C/C++ with kernels executed across GPU threads organized into blocks and grids. CUDA's hierarchy of memories (global, shared, local) optimizes data access patterns for throughput-oriented applications. However, it requires manual memory management (e.g., coalesced accesses) for peak performance, which presents optimization challenges.

Recent advancements in CUDA have focused on enhancing both performance and programmability for HPC applications. One significant development is Unified Memory (UM), which simplifies data management by allowing the CPU and GPU to access the same memory space, reducing the need for explicit data transfers in many cases. NVIDIA also provides a rich ecosystem of CUDA libraries, such as cuBLAS for linear algebra, cuFFT for fast Fourier transforms, and NCCL for multi-GPU communication, which are highly optimized for NVIDIA hardware and are crucial for achieving high performance in areas like scientific computing and deep learning. NVIDIA continuously releases updates to the CUDA Toolkit, introducing new features and optimizations that take advantage of the latest advancements in GPU architectures.

In large-scale GPU-accelerated HPC environments, managing CUDA contexts and data transfers efficiently presents significant challenges. The movement of data between the CPU and GPU memory can often become a bottleneck, limiting the overall performance of applications [1]. To maximize GPU performance, programmers must focus on efficient memory management techniques, including ensuring coalesced memory access patterns, where threads in a warp access contiguous memory location, and effectively utilizing shared memory, a fast on-chip memory that can be shared among threads within a block. In very large applications, careful management of CUDA contexts is also necessary to avoid performance overhead associated with frequent context switching.

Best practices for developing high-performance CUDA applications include overlapping data transfers with kernel execution to hide latency and optimizing the configuration of kernel launches to maximize GPU utilization. In distributed HPC systems with multiple GPUs across multiple nodes, hybrid programming models that combine CUDA for GPU-level parallelism with MPI for inter-node communication are often essential for achieving scalability.

Key strengths of CUDA include:
1) Exceptional performance for SIMD (Single Instruction, Multiple Data) workloads.
2) Direct hardware access for fine-grained parallelism.
3) Support for hybrid CPU-GPU execution.
4) Superior throughput for data-parallel tasks.

Key weaknesses include:
5) High communication overhead between CPU and GPU.
6) Steep learning curve for memory management and kernel optimization.
7) GPU errors typically crash the entire application.
8) Specialized hardware requirement (NVIDIA GPUs).

Table I briefly summarizes the three models.

### III. Comparative Performance Evaluation in HPC Applications

Research has extensively compared the performance characteristics of MPI, OpenMP, and CUDA across a diverse range of HPC application domains [1][4][6]. The suitability and efficiency of each programming model often depend on the specific nature of the application, the underlying hardware architecture, and the optimization strategies employed.

*A. Performance and Scalability*

**MPI:** Excels in distributed environments and demonstrates near-linear scalability in distributed setups. For Goldbach's conjecture verification, MPI codes achieved speed-ups of 1.95x (4 to 8 nodes, i.e., 4 to 8 workers/slaves) and 3.72x (4 to 15 nodes) [2]. In experiments on the shortest-path problem, MPI achieved a 24.14s runtime for 1,000-node graph, outperforming MapReduce (14,440s) but lagging behind OpenMP (8.03s) on shared-memory systems [3]. Scalability is constrained by network latency, with performance degrading on clusters lacking high-speed interconnects like InfiniBand.

TABLE I Key Characteristics of MPI, OpenMP and CUDA

| Aspect | MPI (distributed memory) | OpenMP (shared memory) | CUDA (GPU) |
|---|---|---|---|
| Primary Hardware Target | Distributed clusters | Multi-core CPUs | NVIDIA GPUs |
| Parallelism Type | Process-level parallelism | Thread-level parallelism | Massive data parallelism |
| Communication Mechanism | Explicit message passing | Implicit via shared memory | Memory transfers between CPU-GPU |
| Programming Approach | Message passing between processes | Directive-based multithreading (fork-join parallelism) | Extended C/C++ for GPU kernels (hierarchical parallelism) |
| Programming Complexity | High | Low to Medium | Medium to High |
| Strengths | - Scalability across distributed nodes<br>- Flexibility for dynamic data structures<br>- Hybrid compatibility (e.g., OpenMP) | - Low overhead<br>- Ease of integration via directives<br>- Efficient multi-core utilization | - High throughput for data-parallel tasks<br>- Direct GPU hardware access<br>- Hybrid CPU-GPU execution |
| Weaknesses | - High programming complexity<br>- Communication overhead in shared memory<br>- Manual fault tolerance | - Limited to shared memory systems<br>- Load balancing challenges<br>- Minimal fault tolerance | - CPU-GPU communication bottlenecks<br>- Steep learning curve (memory/kernel tuning)<br>- NVIDIA GPU dependency |
| Recent Advancements | MPI-4: Partitioned communications, Sessions (PMIx), Optimized libraries (Intel MPI, MVAPICH2-GDR) for GPUs/networks | OpenMP 5.0+/6.0: GPU offloading, task enhancements, Improved data management (meta-directives, Fortran/C++ support) | Unified Memory, optimized libraries (cuBLAS, cuFFT, NCCL), Enhanced CUDA Toolkit for latest GPU architectures |

**OpenMP:** Demonstrates superior performance on shared-memory architectures. For image segmentation, OpenMP reduced execution time by 2.5x compared to serial code [6], whereas MPI showed inconsistent scaling with increased processes. OpenMP scales effectively on multi-core CPUs but is limited by shared memory contention. Hyper-Threading provided 41-43% gains over physical cores [2].

**CUDA:** Dominates data-parallel tasks. Matrix multiplication (5,120×5,120 matrix size) completed in 6 seconds on an NVIDIA Tesla K40 GPU, versus 53 minutes for serial CPU code [1]. However, CPU-GPU communication overhead can negate gains for small datasets. CUDA excels in compute-bound tasks (e.g., matrix operations), but performance varies with kernel optimization. Suboptimal implementations lagged behind MPI/OpenMP by orders of magnitude.

*B. Scientific Simulations*

In the realm of scientific simulations, hybrid parallelization strategies have gained significant traction. For instance, in Computational Fluid Dynamics (CFD) applications, a hybrid approach combining MPI for inter-node communication and CUDA for GPU-based computation within each node has demonstrated substantial speedups compared to traditional CPU-based parallel computing. Studies have shown that multi-GPU parallelization using this hybrid model can achieve speedups exceeding 36 times relative to CPU-based parallel implementations [8], highlighting the potential of leveraging GPU acceleration for complex simulations.

For molecular dynamics (MD) simulations, hybrid MPI+OpenMP approaches have proven effective [9], especially on clusters of multi-core nodes, offering a way to balance memory usage and communication overhead.

In weather forecasting models, performance comparisons indicate the significant benefits of GPU acceleration using CUDA for certain computational kernels, while hybrid MPI+OpenMP or MPI+CUDA models are often employed for large-scale simulations on HPC clusters, effectively managing both inter-node and intra-node parallelism.

*C. Machine Learning and Data Analytics*

In the domain of machine learning ML (and deep learning DL), CUDA has become a dominant force for accelerating training and inference tasks on GPUs due to its ability to exploit massive parallelism [10]. For training large models across multiple nodes in an HPC infrastructure, MPI is often employed to handle the distributed communication aspects [7]. OpenMP can also contribute to parallelizing machine learning algorithms, particularly on multi-core CPUs. Performance comparisons often indicate that CUDA can outperform OpenMP for GPU-accelerated machine learning tasks. Hybrid approaches combining MPI and CUDA are frequently adopted for large-scale distributed machine learning to leverage the strengths of both inter-node and intra-node parallelism.

In the realm of data analytics [4], MPI is recognized for its suitability for data-intensive HPC applications requiring portability and scalability across distributed memory systems. OpenMP can provide acceleration for data analytics tasks on shared memory systems by parallelizing loops and other computationally intensive sections of code. CUDA can also be employed to accelerate specific data processing steps on GPUs, offering significant performance gains for certain types of analytical tasks. Hybrid MPI+OpenMP approaches are also prevalent in parallelizing data analytics applications on HPC clusters.

*D. Application Domain Suitability*

The choice of the most suitable programming model is significantly influenced by the inherent characteristics of the application, such as its communication patterns and data dependencies:

**MPI:** Ideal for distributed, communication-heavy applications (e.g., climate modeling, fluid dynamics) requiring inter-node communication.

**OpenMP:** Best suited for shared-memory, loop-centric tasks (e.g., image processing, matrix operations) with minimal synchronization.

**CUDA:** Optimal for data-parallel workloads (e.g., deep learning, signal processing, image processing) leveraging GPU throughput.

Applications that involve substantial communication between different parts of the program, especially across multiple computing nodes, are typically well-suited for MPI, which is designed for distributed memory systems. For applications that exhibit loop-level parallelism and are executed on shared memory systems, OpenMP can provide an efficient and relatively straightforward way to achieve parallel execution. In contrast, applications with a high degree of data parallelism, where the same operation is performed on many data elements simultaneously, often find CUDA's many-core architecture on GPUs particularly advantageous.

*E. Ease of Use and Development Effort*

**OpenMP** leads in usability, with studies rating it higher than MPI and CUDA [2]. Its directive-based model reduces code length and accelerates development. The simplicity of adding pragmas to existing code allows for incremental parallelization without major restructuring.

**MPI** requires explicit handling of data distribution and synchronization, increasing code complexity. For example, MPI's parallel I/O routines struggle with files exceeding 2GB [4]. The need to manually manage communication patterns and data partitioning increases development time and error potential.

**CUDA** demands expertise in GPU architecture. Kernel development and memory optimization (e.g., shared memory usage) are critical for performance, often requiring significant code restructuring. Its balance of abstraction with control comes at the cost of a steep learning curve, particularly for memory management optimizations like coalesced access patterns.

Table 2 provides a brief summary of the section III: *comparative performance evaluation*.

**IV. The Synergy of Hybrid Programming Models**

Hybrid programming models have emerged as a powerful strategy in HPC, seeking to harness the complementary strengths of individual parallel programming approaches [5]. By combining models like MPI, OpenMP, and CUDA, developers can tailor their parallelization strategy to different levels of the computing hierarchy in modern HPC systems. These hybrid models are essential for addressing the heterogeneity found in contemporary HPC architectures.

Table II MPI, OpenMP, and CUDA Performance Comparison Across Different Domains

| Aspect | MPI | OpenMP | CUDA |
|---|---|---|---|
| Performance & Scalability | - Near-linear scalability in distributed setups (e.g., 3.72x speedup for Goldbach's conjecture)<br>- Constrained by network latency (24.14s for 1k-node shortest-path) | - Superior on shared-memory<br>- outperformed MPI on 1k-node shortest-path (8s)<br>- Hyper-Threading gains (41-43% over physical cores) | - Dominates data-parallel tasks (5,120×5,120 matrix in 6s vs. 53min CPU)<br>- Bottlenecked by CPU-GPU transfers for small datasets |
| Scientific Simulations | - Hybrid MPI+CUDA in CFD: 36x speedup over CPU<br>- MPI+OpenMP also effective for molecular dynamics (MD) | - Used in hybrid MPI+OpenMP for memory-efficient MD simulations | - Key for GPU-accelerated CFD/weather models<br>- Hybrid MPI+CUDA manages inter/intra-node parallelism |
| Machine Learning (ML/DL) | - Manages distributed communication in multi-node training | - Parallelizes ML algorithms on multi-core CPUs | - Dominates training/inference (NVIDIA libraries)<br>- Hybrid MPI+CUDA for large-scale distributed ML |
| Domain Suitability | - Communication-heavy apps (climate modeling, fluid dynamics) | - Loop-centric, shared-memory tasks (image processing, matrix ops) | - Data-parallel workloads (DL, image/signal processing) |
| Ease of Use | - High complexity: explicit data/sync management<br>- Challenging I/O for large files (>2GB) | - Simplest: directive-based, minimal code changes<br>- Rated highest usability in studies | - Steep learning curve (GPU memory/kernel tuning)<br>- Requires architectural expertise |

*A. MPI + OpenMP*

One of the most common hybrid approaches involves the combination of MPI and OpenMP. In this model, MPI typically handles the communication and data distribution across multiple nodes in a distributed memory system, while OpenMP manages the parallel execution within each node on the shared memory multi-core processors [2]. This two-level communication pattern intuitively matches the architecture of many HPC clusters, where nodes are interconnected via a network, and each node consists of multiple cores sharing memory.

Studies have demonstrated the performance benefits of this hybrid MPI+OpenMP approach for various applications, including molecular dynamics and scientific simulations [9], often outperforming pure MPI or pure OpenMP implementations, especially on clusters of multi-core nodes.

The combination effectively balances inter-node communication (MPI) with intra-node threading (OpenMP), leading to more efficient resource utilization.

*B. MPI + CUDA*

Another increasingly popular hybrid model combines MPI with CUDA to leverage the power of GPUs in distributed computing environments. In this approach, MPI is used for data distribution and communication between GPU nodes, while CUDA is employed as the main execution engine on each GPU.

Studies have shown that this hybrid MPI/CUDA model can lead to significant acceleration in applications like CFD [8], where computationally intensive tasks can be offloaded to GPUs, and MPI ensures scalability across multiple GPUs and nodes.

*C. OpenMP + CUDA*

This hybrid approach leverages CPU cores and GPU threads concurrently, combining the ease of use of OpenMP for CPU parallelization with the massive parallelism of CUDA for GPU acceleration. However, synchronization overheads can diminish gains without careful load balancing [11]. In matrix multiplication, this hybrid model achieved speedup over standalone CUDA.

*D. MPI + OpenMP + CUDA*

More complex hybrid models that combine all three programming paradigms—MPI, OpenMP, and CUDA—are also being explored to further enhance performance by leveraging the strengths of each model at different levels of parallelism [11]. These models can offer enhanced performance and scalability by strategically distributing the workload across different types of processing units.

*E. Implementation Challenges*

Designing and implementing efficient hybrid parallel applications requires careful consideration of several factors [11]:
1) Workload decomposition and distribution: Effective partitioning of tasks across the inter-node (MPI) and intra-node (OpenMP or CUDA) levels of parallelism is crucial for maximizing resource utilization.
2) Synchronization management: Coordinating execution across different programming models requires careful attention to synchronization points to avoid deadlocks and race conditions.
3) Communication optimization: Minimizing communication overhead between MPI processes and efficiently managing data transfers within shared memory regions or between the CPU and GPU are critical aspects.
4) Load balancing: Achieving proper load balancing across the different processing units, whether they are CPU cores or GPUs, requires careful planning and often dynamic adjustment during runtime.

5) Debugging complexity: Hybrid models introduce additional layers of complexity in debugging and performance analysis, requiring sophisticated tools and methodologies.

Despite these challenges, hybrid programming models offer a promising approach to fully exploit the computational potential of modern heterogeneous HPC systems, and their adoption is likely to continue growing as hardware architectures become increasingly diverse.

## V. Challenges and Best Practices

Deploying and scaling parallel applications using MPI, OpenMP, and CUDA on large-scale HPC systems and emerging architectures presents a unique set of challenges. Modern HPC systems are increasingly characterized by heterogeneous architectures [11], featuring a mix of distributed memory across nodes, shared memory within nodes, and various types of accelerators, which adds complexity to parallel programming.

*A. Scaling Challenges*

**MPI applications** at large scales can encounter limitations due to communication latency and synchronization overhead, which can become significant as the number of processes increases [6]. Ensuring fault tolerance in MPI applications running on thousands of nodes also poses a considerable challenge [11]. MPI's resilience in exascale systems needs enhancement, as current fault tolerance mechanisms may not scale effectively.

**OpenMP applications**, while simplifying shared memory parallelization, can face scalability issues on very large core counts hindered by shared memory contention and the difficulty of achieving effective load balancing across all cores [11]. Fine-grained synchronization can become a bottleneck as thread counts increase.

**CUDA applications** running on large GPU clusters face challenges in managing the distribution of data across multiple GPUs, ensuring efficient inter-GPU communication, and effectively utilizing the GPU memory hierarchy [11]. Also, the high communication overhead between CPU and GPU can limit performance for applications requiring frequent data exchange.

*B. Best Practices for Optimization*

To address these challenges, implementing best practices [11] is essential:

For MPI:
1) Overlap communication with computation to hide latency.
2) Use non-blocking communication where appropriate.
3) Carefully tune collective communication operations for the specific network topology.
4) Implement topology-aware process placement to minimize communication distances.
5) Utilize one-sided communication operations where applicable to reduce synchronization overhead.

For OpenMP:
1) Apply judicious loop scheduling to ensure balanced workload distribution.
2) Manage data locality effectively to maximize cache utilization.
3) Leverage advanced features like tasking for irregular parallelism.
4) Use thread affinity to bind threads to specific cores for improved cache performance.
5) Minimize synchronization points and use fine-grained synchronization constructs.

For CUDA:
1) Maximize GPU occupancy to fully utilize available resources.
2) Ensure coalesced access to global memory for improved memory bandwidth.
3) Minimize data transfers between the CPU and GPU.
4) Use asynchronous operations to overlap computation and data transfers.
5) Leverage shared memory and register usage for frequently accessed data.

For hybrid models:
1) Balance the workload and communication responsibilities between the different programming paradigms.
2) Allocate resources appropriately between MPI processes and OpenMP threads.
3) Optimize data locality to minimize cross-node and cross-core communication.
4) Implement dynamic load balancing for heterogeneous workloads.
5) Use profiling tools to identify and address performance bottlenecks.

*C. Emerging Architecture Considerations*

The emergence of new architectures introduces additional considerations:

**GPU-accelerated Arm-based systems** [12] introduce complexities related to porting and optimizing applications for these platforms. The combination of Arm CPU architecture with GPU accelerators requires careful tuning of both CPU and GPU code.

**Intel Xeon Phi** [5] competes with GPUs by offering many-core CPUs with AVX-512 vectorization. While CUDA requires code porting, OpenMP programs run natively on Xeon Phi with minimal changes.

**Non-Volatile RAM (NVRAM)** [7] has the potential to revolutionize checkpointing and I/O-heavy workflows by providing persistent memory that bridges the gap between volatile main memory and storage.

**Energy efficiency** considerations are becoming increasingly important in HPC [7]. Power-aware configurations (e.g., GPU power capping) are critical for sustainable HPC, though current APIs lack built-in mechanisms for energy-aware scheduling.

## VI. Future Trends and Evolving Landscape of Parallel Programming in HPC

The landscape of parallel programming in HPC is continuously evolving, driven by advancements in hardware architectures and the increasing demands of scientific applications. Several key trends are shaping the future of how HPC software is developed and executed.

### A. Evolution of Individual Programming Models

**MPI Evolution:** Future directions for MPI include further enhancements to support hybrid programming models more effectively and better integration with accelerators. This involves developing more sophisticated communication primitives that can efficiently handle the complexities of heterogeneous systems and provide greater flexibility for overlapping communication and computation. The ongoing development of MPI implementations that are "CUDA-aware" or that integrate effectively with other accelerator technologies represents a significant area of advancement.

**OpenMP Advancements:** OpenMP is undergoing significant advancements, particularly in expanding its support for heterogeneous architectures, including GPUs and other accelerators. Future developments are likely to focus on further improving the usability and performance of task-based parallelism, making OpenMP a more comprehensive solution for a wider range of HPC platforms, potentially extending beyond traditional shared memory systems.

**CUDA Development:** CUDA is expected to maintain its strong position as the leading platform for GPU computing in HPC. Future directions will likely involve continued performance optimizations for existing and emerging NVIDIA GPU architectures, as well as the development of new features and libraries tailored to the evolving needs of AI and scientific computing workloads. CUDA's ecosystem of specialized libraries for various domains continues to expand, providing highly optimized implementations for common HPC tasks.

### B. Emerging Programming Models and Frameworks

**Performance Portability Frameworks:** There is a growing interest in performance portability frameworks like OpenACC, SYCL, and Kokkos [7]. These frameworks aim to provide a level of abstraction that allows developers to write parallel code that can run efficiently on a variety of different hardware architectures without requiring extensive modifications. As HPC systems become more heterogeneous, the need for such frameworks will likely increase to simplify development and ensure that applications can effectively utilize the available computing resources. Frameworks like Kokkos and RAJA simplify cross-platform development but remain underutilized in many HPC environments [7].

**Task-Based Programming:** Task-based programming models are gaining attention as an alternative to traditional data parallelism [13]. These models allow developers to express applications as a set of tasks with dependencies, leaving the runtime system to schedule and execute them efficiently. This approach can potentially lead to better load balancing and resource utilization in heterogeneous environments.

**Partitioned Global Address Space (PGAS):** PGAS models provide a global view of memory across distributed systems, simplifying programming while maintaining awareness of data locality [14]. Languages and libraries implementing the PGAS concept, such as Chapel, UPC++, and Coarray Fortran, are seeing increased adoption in certain HPC domains.

### C. HPC-Big Data Convergence

Frameworks like Apache Spark are increasingly being integrated with MPI for iterative algorithms, bridging data-intensive and compute-intensive workflows [4][7]. This convergence of HPC and Big Data paradigms is driven by the need to process and analyze massive scientific datasets efficiently.

Spark's Resilient Distributed Datasets (RDDs) are inspiring fault-tolerant HPC models [4], though integration with MPI/OpenMP remains nascent. The ability to recover from failures without restarting the entire application is becoming increasingly important as systems scale to exascale levels.

### D. Architectural Innovations

The ongoing architectural innovations in HPC, including the advent of exascale systems and the integration of specialized hardware like AI accelerators and potentially quantum processors [15] in the future, will continue to drive the evolution of parallel programming models and software development practices in the field.

Energy efficiency is becoming a critical consideration in HPC system design [7]. Future programming models may need to incorporate energy-aware constructs to allow developers to make trade-offs between performance and power consumption.

These trends collectively point towards a future where parallel programming in HPC becomes both more powerful and more complex, requiring sophisticated tools, frameworks, and methodologies to effectively harness the computational capabilities of emerging architectures.

## VII. Conclusions

The comparative study of MPI, OpenMP, and CUDA reveals a complex landscape of parallel programming for HPC applications. Each model offers distinct advantages and is best suited for specific types of hardware and computational tasks. MPI remains the cornerstone for distributed memory parallelism, essential for scaling applications across the nodes of HPC clusters. Its ongoing evolution addresses the challenges posed by modern architectures, with a focus on improved integration with multi-threading and accelerators. OpenMP has expanded its capabilities beyond shared memory CPUs to include support for accelerators, offering a higher-level, directive-based approach to parallelization on heterogeneous systems. Its continuous development aims to enhance performance and programmability, making it a versatile option for HPC. CUDA continues to be the dominant force in GPU-accelerated computing, providing a powerful platform for massively parallel computations, particularly in domains like AI. The rich ecosystem of CUDA libraries and the continuous introduction of new features ensure its sustained importance in HPC.

The emergence and increasing adoption of hybrid programming models, combining MPI with OpenMP or CUDA, or even all three, signify a crucial trend in addressing the complexities of modern HPC systems. These hybrid approaches allow developers to leverage the strengths of each model at different levels of parallelism, leading to improved performance and scalability for a wider range of applications. Furthermore, the growing interest in performance portability frameworks underscores the need for solutions that can effectively manage the increasing heterogeneity of HPC architectures, enabling code to run efficiently across diverse computing resources without extensive modifications.

Based on our analysis, we offer the following key recommendations:
1) Use MPI for distributed, communication-bound applications requiring inter-node parallelism.
2) Prefer OpenMP for shared-memory, loop-parallel tasks with minimal synchronization requirements.
3) Deploy CUDA for data-parallel workloads with large datasets that can benefit from GPU acceleration.
4) Explore hybrid models to optimize resource utilization in heterogeneous environments.

Future work should focus on standardizing fault tolerance mechanisms for large-scale systems, enhancing interoperability between HPC and Big Data ecosystems, and developing programming models that can effectively address the challenges of exascale computing and emerging heterogeneous architectures.

In conclusion, the choice of parallel programming model, or combination thereof, for HPC applications is a critical decision that depends on a multitude of factors, including the application's inherent parallelism, communication requirements, the target hardware architecture, and the desired level of developer productivity. The evolving landscape of HPC suggests a future where a combination of these models, along with emerging paradigms and architectural innovations, will be key to unlocking the full potential of high-performance computing for scientific discovery and technological advancement.